# Searches for Stable Strangelets in Ordinary Matter: Overview and a Recent Example

Z.-T. Lu [*a], R. J. Holt [a], P. Müller [a], T. P. O'Connor [a], J. P. Schiffer [a,b], L.-B. Wang [a,c]

[a] *Physics Division, Argonne National Laboratory, Argonne, Illinois 60439, USA*
[b] *Physics Department, University of Chicago, Chicago, Illinois 60637, USA*
[c] *Physics Department, University of Illinois at Urbana-Champaign, Urbana, Illinois 61801, USA*

**Abstract**

Our knowledge on the possible existence in nature of stable exotic particles depends solely upon experimental observation. Guided by this general principle and motivated by theoretical hypotheses on the existence of stable particles of strange quark matter, a variety of experimental searches have been performed. We provide an introduction to the theoretical hypotheses, an overview of the past searches, and a more detailed description of a recent search for helium-like strangelets in the Earth's atmosphere using a sensitive laser spectroscopy method.

Is it possible that ordinary nuclear matter that we are familiar with is actually metastable and slowly decaying into quark matter of lower energy? Many hypotheses have been proposed addressing this question. Bodmer first proposed a lower-energy nuclear state, which he called "collapsed nuclei" [1]. Witten investigated the stability of strange quark matter, which consists of approximately equal number of *u, d* and *s* quarks in a hadronic state, and reached the intriguing conclusion that it may indeed be stable even at zero external pressure [2]. A particle of strange quark matter is called a "strangelet". Searches for such exotic matter have been performed in astronomical observation, in cosmic rays, in heavy-ion collisions, and in chemical analyses of ordinary matter. Several excellent review articles on this subject are available [3-5]. Here we focus on one aspect of the subject: the experimental searches for stable strangelets in ordinary matter. In Section 1, we provide a brief pedagogical introduction to strange quark matter and strangelets; in Section 2, we review the past searches; in Section 3, we present our recent search for helium-like strangelets in the Earth's atmosphere.

## 1. Introduction

Ordinary nuclear matter consists of *u* and *d* quarks confined in nucleons. Since quarks are fermions, the Fermi energy of a nucleus could be lowered by introducing the extra degree of freedom of *s* quarks. On the other hand, because an *s* quark ($\approx 150$ MeV/c$^2$) is more massive than a *u* or *d* quark ($\approx 4$ MeV/c$^2$), extra energy is needed to generate s quarks. The intricate balance of these two opposing factors determines the stability of strange quark matter. Such low-energy problems cannot yet be solved with QCD theory from first principles; instead, model-dependent calculations have been performed. Jaffe *et al.* found that in order to obtain stable strange matter it was necessary to reduce the value of the "bag constant" below traditionally favored value [4]. They concluded that the stability of strange quark matter at zero external pressure is unlikely, but possible.

---

[*] Email: lu@anl.gov

If strange matter has lower energy, the decay of ordinary matter via weak interaction is expected to be extremely slow and difficult to observe directly; it should be much slower than double beta decays, which typically have lifetimes on the order of $10^{20}$ years. Strange matter is likely to be produced in the core of neutron stars where the extremely high pressure convert ordinary matter into strange matter. Strangelets would then be released into space in such violent events as the collision of two stars, and could become part of the terrestrial matter.

A stable strangelet would consist of roughly equal numbers of *u, d*, and *s* quarks with somewhat fewer *s* quarks due to their higher mass. Such a strangelet would be positively charged with a much lower charge-over-mass ratio than an ordinary nucleus, and be surrounded by electrons to form an atom with an exotic core. In chemical analyses, it would appear as an anomalously heavy isotope of an ordinary element. For example, a strangelet of +2e charge would combine with two electrons to form an anomalously heavy helium atom. Note that its mass doesn't have to be an integer multiple of the nucleon mass.

**2. Past searches**

Many methods have been employed to search for stable strangelets in a wide variety of samples of ordinary matter; none have been found. These searches cover not only strangelets, but also other exotic particles that would produce similar effects probed in these experiments. Figure 1 shows the upper limits on the abundance of strangelets set by past searches. Although it is convenient to present these limits in one figure, we caution that there is only a weak relationship between the limits set by these searches, which are based on very different sets of assumptions. For example, by far the most stringent limits have been set on +1e charged, hydrogen-like strangelets [6]. However, the non-existence of +1e charged strangelets does not automatically rule out the existence of strangelets of higher charges. Moreover, each experiment is based on some specific assumptions, e.g. on the distribution of strangelets in the environment, and on the physical and chemical behavior of strangelets. An important issue that all strangelet hunters must address is what samples to use and where to collect them. Some examples are given in the following discussion.

**2.1 Mass spectrometry (MS), $^2$H**

Smith *et al.* [6] took advantage of the existing industrial scale heavy-water enrichment facility. Starting from $1.2 \times 10^8$ l of natural water, $2 \times 10^{-5}$ l of D$_2$O sample was extracted with multi-step electrolysis, first in a factory and then in the laboratory. Based on the theory of electrolysis, the authors concluded that 10% of the H-like strangelets in the initial natural water sample would remain throughout processing and the density of strangelets enriched by an overall factor of $6 \times 10^{11}$. The enriched sample was then injected into a novel mass spectrometer specifically designed to detect singly charged massive particles. No event of abnormally heavy H was observed, from which an abundance limit of X/D ~ $10^{-16}$–$10^{-17}$ in the sample was established and, when combined with the enrichment factor, a limit of X/H ~ $10^{-28}$ – $10^{-29}$ over the mass range of 12–1200 amu was obtained.

Interestingly, simply by verifying that the density of the commercial $D_2O$ and that of the further enriched $D_2O$ are equal within 10%, a new limit X/H ~ $10^{-14}$–$10^{-17}$ over the mass range of $10^3$–$10^6$ amu was established. The large enrichment factor was again factored into the limits, although it is not clear whether the theory of electrolysis is applicable in this higher mass range. In addition, it is difficult to estimate whether very heavy water molecules would remain part of the circulating water reservoir that was sampled.

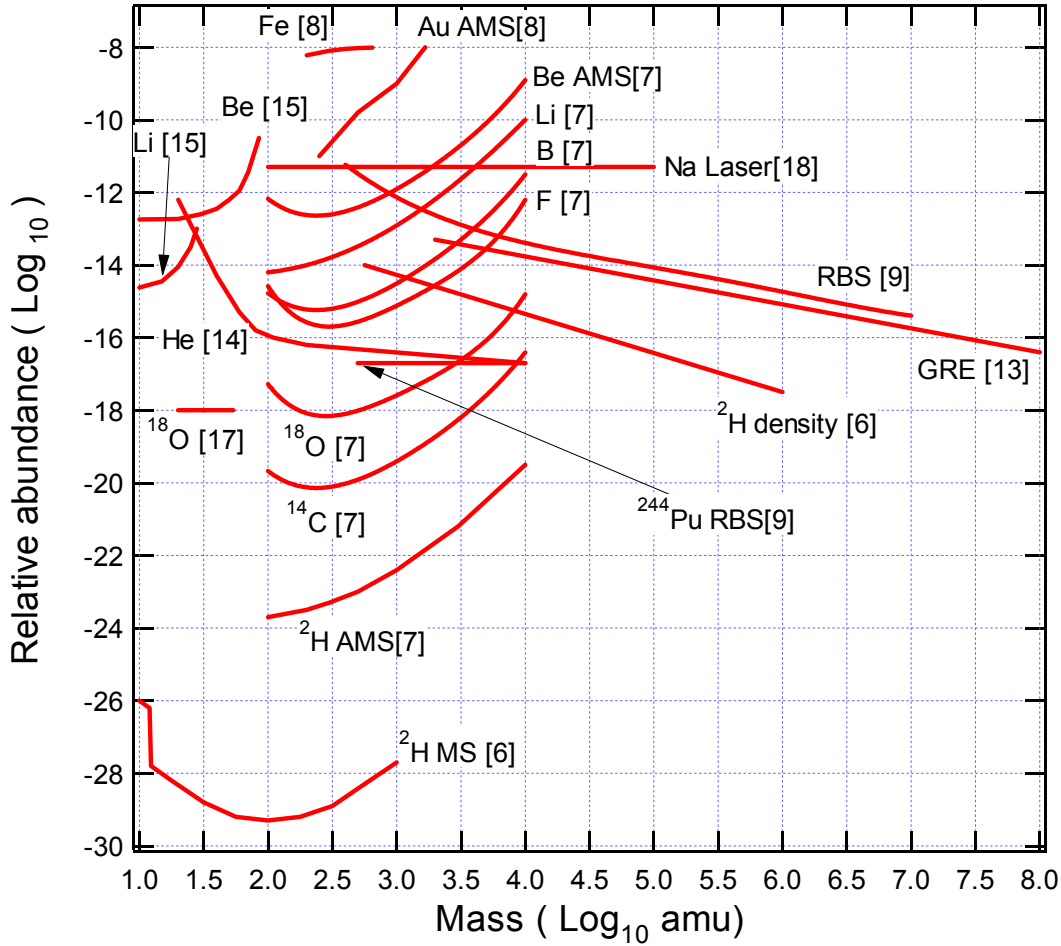

**Figure 1.** Published experimental upper limits on the abundance of strangelets. Reference numbers are indicated in brackets. We caution that there is only a weak relationship between the limits set by these searches, which are based on very different sets of assumptions. The methods used are: mass spectrometry [6]; accelerator mass spectrometry [7, 8, 15, 17]; density [6]; laser spectroscopy [14, 18]; Rutherford back scattering [9]; γ-ray emission [13].

### 2.2 Accelerator mass spectrometry (AMS), $^2$H, Be, Li, B, F, $^{18}$O, $^{14}$C

The low background achieved in the previous low-energy (10-100 keV) mass analysis is only feasible with +1e charged particle due to the complete absence of mass isobars. For higher charges, both atomic and molecular isobar contamination would greatly reduce the sensitivity of such searches. Isobaric contamination could be efficiently eliminated in some cases by performing mass spectrometry with a high-energy (~ MeV) beam produced by an accelerator. First, some atomic isobars can be expelled by exploiting the stability property of negative ions that are used in the first acceleration stage of a tandem accelerator. For example, $^{14}N^-$, the only abundant atomic isobar of $^{14}C^-$, is not stable, and consequently is not produced or accelerated. Second, molecular isobars can be efficiently removed by passing the accelerated beam through a thin foil that strips away some electrons from the molecule and causes the molecule to disintegrate. Third, with high-energy (~ MeV) ions, more versatile and discriminatory ion detection techniques such as the energy-loss measurement can be applied to help identify the core charge of the ions and further reduce the influence of isobaric interferences.

Hemmick *et al*. [7] used the AMS spectrometer at the University of Rochester to search for anomalously heavy isotopes and set abundance limits of $10^{-15}$–$10^{-10}$ (before including the enrichment factors) in the mass range of $10^2$–$10^4$ amu. In this work, the samples of Li, Be, B, and F were obtained from common commercial chemicals. The $^2H$ samples were enriched from natural water by a factor of $10^9$, the $^{14}C$ sample was enriched by a factor of $10^5$, and $^{18}O$ by a factor of $10^2$.

Javorsek *et al*. [8] used the AMS spectrometer at the PRIME Lab of Purdue University to search for heavy isotopes of Au and Fe and set abundance limits of X/Au ~ $10^{-11}$–$10^{-8}$ and X/Fe ~ $10^{-9}$ in the mass range of $10^2$–$10^3$ amu. For this search, Au samples include near-surface geological samples, a piece of a satellite, and a piece of the RHIC beam dump; the Fe sample was taken from a meteorite.

**2.3 Rutherford back scattering (RBS)**

The back scattering of a projectile nucleus in a single collision, a phenomenon first employed by Rutherford to investigate the nature of atomic nuclei, can only be induced by a nucleus or particle in the target that is heavier than the projectile nucleus. At GSI in Germany [9], the back scattering of a $^{238}U$ beam was measured to search for strangelets of any charge in the mass range of $4 \times 10^4$–$10^7$ amu. The energy of the beam (1.4 MeV per nucleon) was kept far below the coulomb barrier to ensure that the collisions were elastic. The upper limit of the mass range was set at $10^7$ amu due to the concern over the screening of the nuclear charge by electrons inside or in close proximity of large strangelets. An iron meteorite and various types of terrestrial samples were used as targets. One notable advantage of RBS over MS or AMS is that its sample processing is rather straightforward: the samples need not be purified, nor do they need to be vaporized or ionized as in the case of MS or AMS.

Polikanov *et al*. [9] used RBS to search for strangelets in a $^{244}Pu$ ($t_{1/2} = 8 \times 10^7$ yr) target. Since Pu-like strangelets may be stable, over the age of the Earth the strangelets would be enriched by a factor of $10^{17}$ as the ordinary Pu decay. This enrichment scheme also applies to other elements with no stable isotopes, such as Tc, Pm and Rn.

**2.4 Gamma-ray emission (GRE)**

Strangelets can grow in mass by absorbing ordinary nuclear matter and, in the process, release the excess energy in the form of photons and particles. In this conference, Bombaci presented a theoretical investigation on the process of strangelet growth in a neutron star [10]. If strangelets are stable at zero pressure, a strangelet in a neutron star will keep growing until the whole star is converted into a gigantic strangelet, or a strange matter star. Bombaci postulated that this conversion process can be the physical source of the observed giant γ-ray emission events.

This reaction process also presents another way of searching for strangelets in laboratories. Holt *et al*. [11] measured the γ-rays in an anomalously high energy range (30–250 MeV) when a sample is irradiated by a thermal neutron beam generated by the CP-5 nuclear reactor at Argonne National Laboratory. The search was focused on Rn-like strangelets because the primordial noble gas on Earth is concentrated in the atmosphere, and in order to take advantage of the enrichment factor enjoyed by elements with no stable isotopes. They concluded that the weight of such strangelets on the Earth is $< 10^{-29}$ of the total weight of the Earth, and the relative weight in the solar system is $< 10^{-24}$.

Compared with neutron capture, the released energy would be multiplied by the mass number of a nucleus that penetrates the Coulomb barrier of the strangelet, thus making heavy ion activation an attractive search method. Farhi and Jaffe [12] proposed that the excess energy (~ GeV) can be released in the form of a large number of low-energy ($10-10^3$ keV) photons. Perillo-Isaac *et al*. [13] carried out such a search with a $^{136}$Xe beam generated by the 88" Cyclotron at Lawrence Berkeley National Laboratory. They employed the Gammasphere detector to search for the signature of multi-photon emission events. Samples included nickel ore, Allended meteorite, and lunar soil. The upper limits on abundance X/(sample nucleons) are $10^{-13}$–$10^{-17}$ in the mass range $10^3$–$10^8$ amu. Above $10^8$ amu the photons were expected to have energy (< 20 keV) too low to be detected with Gammasphere.

## 3. Laser spectroscopic search for He-like strangelets in the Earth's atmosphere

In this section, we discuss a recent search for He-like strangelets in the Earth's atmosphere using a laser spectroscopy method [14]. The He-like case is particularly favorable because ordinary helium is severely depleted in the terrestrial environment. Due to its low mass, ordinary helium (mostly $^4$He) is likely to be depleted in the formation of the Earth. Any remaining ordinary helium finds its way to the exosphere, from where it escapes into space, and thus helium is replenished only from radioactive decay. Other noble gases, from neon to xenon, are concentrated largely in the Earth's atmosphere, after an initial, lesser depletion relative to solar system levels at the early stages of the planet's evolution. The concentration of noble-gas-like atoms in the atmosphere and the subsequent very large depletion of the known light $^{3,4}$He isotopes from the atmosphere allow significantly enhanced limits to be set.

Searches specifically aimed at anomalous isotopes of helium have been performed in two previous experiments. With AMS, Klein *et al.* [15] set an isotopic abundance limit of $6 \times 10^{-15}$ over the mass range of 3–8 amu. With MS, Vandegriff *et al.* [16] set an isotopic abundance limit of $10^{-3}$–$10^{-5}$ over the mass range of 42–82 amu.

The laser spectroscopy method took advantage of the isotope shift due to the higher mass of a heavier nucleus. In helium, the isotope shift is dominated by the mass shift that results from the change of the nuclear mass. The mass shift $\delta v_{MS}$ of a transition between isotopes of nuclear mass $M$ and mass infinity is given as $\delta v_{MS} = -F_{MS}/M$. This technique is particularly suited to searches for isotopes with unknown mass because the range of atomic transition frequency to be searched is finite even as the atomic mass goes to infinity. We performed the search by probing the 1s2s $^3$S$_1$ → 1s2p $^3$P$_2$ transition at 1083 nm in helium atoms at the metastable 1s2s $^3$S$_1$ level. From the known $^3$He - $^4$He isotope shift, it is derived that $F_{MS}$ = 412 GHz for this transition.

In order to search for a weak absorption signal and avoid Doppler broadening, we performed frequency-modulation saturation spectroscopy on the 1s2s $^3S_1$ → 1s2p $^3P_2$ transition (natural linewidth = 1.6 MHz). The helium sample was extracted from air with sorption pumps cooled to 80 K by liquid nitrogen, thus effectively absorbing all major gases in air except neon and helium. The sample was then transported into a long glass cell for spectroscopic analysis. The calibration of the detection sensitivity was accomplished by using $^3$He as a reference isotope, whose abundance in air is $1.4 \times 10^{-6}$ relative to $^4$He. The observed signal of $^3$He in the atmospheric sample had a signal-to-noise ratio of 70 and a linewidth of 45 MHz. The $^3$He lineshape was important, as it was used as a template when looking for a signal. For the search the laser was slowly scanned across a range of mass shift (relative to He of mass infinity) $\delta\nu_{MS}$ from −86.3 to +3.1 GHz, corresponding to a mass range of 4.8 amu – infinity. The data were used to search for a possible signal by sliding the template over the scan range and calculating the best fit for the amplitude for each frequency with a resolution of 3 MHz. We find the distribution of the fitted amplitudes over the entire frequency range to be statistical with the mean value of zero. We conclude at the 95% confidence level that there is no anomalous peak with an amplitude larger than $7.9 \times 10^{-2}$ times the $^3$He amplitude anywhere in the entire frequency range. For masses above this value, the shrinking Doppler width may in fact alter the signal linewidth and reduce the detection sensitivity. For this reason we conservatively set the high-mass limit at $1 \times 10^4$ amu.

The origin and evolution of the terrestrial noble gases have been subjects of great interest in geology for the past five decades. Although it is not conclusively established, past geochemical measurements indicate and most global evolution models infer or assume, that the total inventory of primordial noble gases in the mantle is small compared to that in the atmosphere. For the case of helium, it is known that the $^{3,4}$He nuclei present in the Earth are almost all "young" nuclei with radiogenic or cosmogenic origin; the lifetime for $^{3,4}$He atoms in the atmosphere is only ~ $2 \times 10^6$ years. Knowing that there are $6 \times 10^{38}$ $^4$He nuclei in the Earth's atmosphere and a total of $1 \times 10^{50}$ nuclei of all kinds in the Earth, we can set limits on the abundance of anomalous helium-like particles in the whole Earth at $10^{-18}$–$10^{-20}$ per atom over the mass range of 5–10,000 amu.

It is believed that the sun and the planets formed from the same starting material, and that this original composition is preserved in the sun. The noble gases, as well as hydrogen, were either not captured in the planet formation process, or were subsequently depleted in the Earth at the early stage when the planet was molten. The deficiency factors for each noble gas element, defined as the ratios of the abundances of the elements in the Earth over those in the Sun, are well documented. There is clearly a mass dependence: heavier noble gas atoms are retained more than the lighter ones, and for atoms of mass > 80 the deficiency factor approaches a constant. Assuming that the deficiency factors for the anomalous helium follow the same mass dependence, we can set limits on their abundance in the solar system at $10^{-12}$–$10^{-17}$ per atom over the mass range of 20–10,000 amu. The sensitivity of the method presented in this paper could be improved further by perhaps several orders of magnitude with the application of cavity-enhanced spectroscopy and by perhaps enriching heavy helium with gas chromatography.

This work is supported by the U.S. Department of Energy, Office of Nuclear Physics, under contract W-31-109-ENG-38.

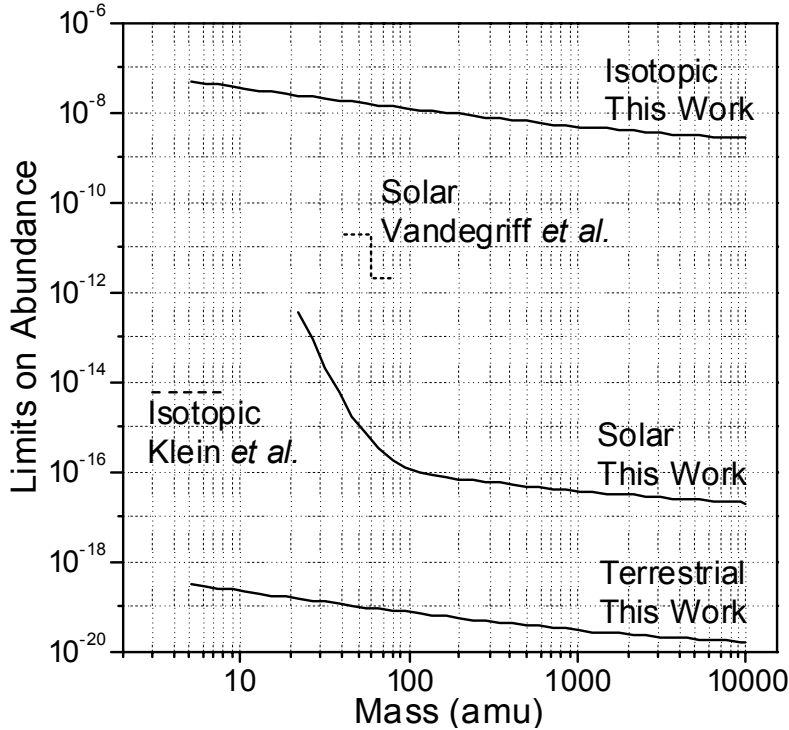

**Figure 2**. Limits on the abundance of anomalously heavy isotopes of helium. The solid lines indicate limits set in this work: top, on the isotopic abundance (anomalous helium vs. $^4$He) in the Earth's atmosphere; middle, on the atomic abundance (anomalous helium vs. total number of nuclei) in the solar system; bottom, on the atomic abundance in the Earth. The left dashed line indicates the limits set by Klein *et al.* [15] on the isotopic abundance in the Earth's atmosphere. The right dashed line indicates the limits set by Vandegriff *et al.* [16] on the atomic abundance in the solar system.